\documentstyle[12pt,epsfig]{article}
\begin{document}
\centerline{\large\bf Time variation of the fine structure constant
in decrumpling}
\vspace*{0.050truein}
\centerline{\large\bf or TVSD model}
\vspace*{0.050truein}
\centerline{Forough Nasseri\footnote{Email: nasseri@fastmail.fm}}
\centerline{\it Physics Department,
Khayyam Planetarium, P.O.Box 769, Neishabour, Iran}
\begin{center}
(\today)
\end{center}

\begin{abstract}
Within the framework of a model universe with time variable space
dimension (TVSD), known as decrumpling or TVSD model, we study the time
variation of the fine structure constant. Using observational bounds
on the present time variation of the fine structure constant, we are
able to obtain an upper limit for the absolute value of the present
time variation of spatial dimensions.
\end{abstract}

\date{today}

Although time variability of spatial dimensions have not been firmly
achieved in experiments and theories, such dynamical behavior of the
spatial dimensions should not be ruled out in the context of cosmology
and astroparticle physics \cite{1}-\cite{7}.

In this letter, we study the time variation of the fine structure
constant in decrumpling or TVSD model.\footnote{It is worth mentioning
that from Eq.(\ref{1}) to Eq.(\ref{19}) we use a natural unit system that
sets $k_B$, $c$ and $\hbar$ all equal to $1$, so
that $\ell_P=M_P^{-1}=\sqrt{G}$. From Eq.(\ref{a20}) to Eq.(\ref{47}) we
use the International System of Units.}

Assume the universe consists of a fixed number ${\bar N}$ of universal
cells having a characteristic length $\delta$ in each of their
dimensions. The volume of the universe at the time $t$ depends
on the configuration of the cells. It is easily seen that
\cite{7}
\begin{equation}
\label{1}
{\rm vol}_{D_t}({\rm cell})={\rm vol}_{D_0}({\rm cell})\delta^{D_t-D_0},
\end{equation}
where the $t$ subscript in $D_t$ means $D$ is as a function
of time.
Interpreting the radius of the universe, $a$, as the radius of
gyration of a crumpled ``universal surface'',
the volume of space can be written \cite{7}
\begin{eqnarray}
\label{2}
a^{D_t}&=&{\bar N} {\rm vol}_{D_t}({\rm cell})\nonumber\\
   &=&{\bar N} {\rm vol}_{D_0}({\rm cell}) \delta^{D_t-D_0}\nonumber\\
   &=&{a_0}^{D_0} \delta^{D_t-D_0}
\end{eqnarray}
or
\begin{equation}
\label{3}
\left( \frac{a}{\delta} \right)^{D_t}=
\left( \frac{a_0}{\delta} \right)^{D_0} = e^C,
\end{equation}
where $C$ is a universal positive constant. Its value has a strong
influence on the dynamics of spacetime, for example on the dimension
of space, say, at the Planck time. Hence, it has physical and cosmological
consequences and may be determined by observations. The zero subscript in any
quantity, e.g. in $a_0$ and $D_0$, denotes its present values.
We coin the above relation as a``dimensional constraint" which relates
the ``scale factor" of the model universe to the space dimension.
In our formulation, we consider the comoving length of the Hubble radius
at present time to be equal to one. So the interpretation of the scale
factor as a physical length is valid.
The dimensional constraint can be written in this form
\begin{equation}
\label{4}
\frac{1}{D_t}=\frac{1}{C}\ln \left( \frac{a}{a_0} \right) + \frac{1}{D_0}.
\end{equation}

It is seen that by expansion of the universe, the space
dimension decreases. 
Time derivative of Eqs.(\ref{3}) or (\ref{4}) leads to

\begin{equation}
\label{5}
{\dot{D}}_t=-\frac{D_t^2 \dot{a}}{Ca}.
\end{equation}
It can be easily shown that the case of constant space dimension
corresponds to when $C$ tends to infinity. In other words,
$C$ depends on the number of fundamental cells. For $C \to +\infty$,
the number of cells tends to infinity and $\delta\to 0$.
In this limit, the dependence between the space dimensions and
the radius of the universe is removed, and consequently we
have a constant space dimension.

We define $D_P$ as the space dimension of the universe when the
scale factor is equal to the Planck length $\ell_P$.
Taking $D_0=3$ and the scale of the universe today to be the present
value of the Hubble radius $H_0^{-1}$ and the space dimension at the
Planck length to be $4, 10,$ or $25$, from Kaluza-Klein and superstring
theories, we can obtain from Eqs. (\ref{3}) and (\ref{4})
the corresponding value of $C$ and $\delta$
\begin{eqnarray}
\label{6}
\frac{1}{D_P}&=& \frac{1}{C} \ln \bigg( \frac{\ell_P}{a_0}\bigg) +
\frac{1}{D_0}= \frac{1}{C} \ln \bigg( \frac{\ell_P}{H_0^{-1}}\bigg) +
\frac{1}{3},\\
\label{7}
\delta&=&a_0 e^{-C/D_0}=H_0^{-1} e^{-C/3}.
\end{eqnarray}
In Table 1, values of $C$, $\delta$ and also
${\dot D}_t|_0$ for some interesting values of $D_P$ are
given. These values are calculated by
assuming $D_0=3$ and
$H_0^{-1}=3000 {h_0}^{-1} {\rm Mpc} = 9.2503 \times 10^{27} {h_0}^{-1} 
{\rm cm}$, where $h_0=0.68 \pm 0.15$.
Since the value of $C$ and $\delta$ are not very
sensitive to $h_0$ we take $h_0=1$.

\begin{table}
\caption{Values of $C$ and $\delta$ for some values of
$D_P$ \cite{1}-\cite{7}. Time variation of space dimension today
has also been calculated in terms of 
yr$^{-1}$.}
\begin{tabular}{cccc} \\ \hline\hline 
$D_P$ & $C$   & $\delta$ (cm)   & ${\dot D}_t|_0$ (yr$^{-1}$) \\ \hline\hline
$3$           & $ +\infty$         &  $0$           & $0$ \\ \hline
$4$           & $1678.797$         &  $8.6158 \times 10^{-216}$  & $ -5.4827 \times 10^{-13} h_0$  \\ \hline
$10$          & $599.571$          &  $1.4771 \times 10^{-59}$  & $ -1.5352 \times 10^{-12} h_0$  \\ \hline
$25$          & $476.931$          &  $8.3810 \times 10^{-42}$  & $-1.9299 \times 10^{-12} h_0$ \\ \hline
$+\infty$     & $419.699$          &  $\ell_P$  & $ -2.1931 \times 10^{-12}h_0$ \\ \hline\hline
\end{tabular}
\end{table}

Let us define the action of the model for the special
Friedmann-Robertson-Walker (FRW) metric in an arbitrary fixed space
dimension $D$, and then try to generalize it to variable dimension $D_t$.
Now, take the metric in constant $D+1$ spacetime dimensions in the
following form
\begin{equation}
\label{8}
ds^2 = -N^2(t)dt^2+a^2(t)d\Sigma_k^2,
\end{equation}
where $N(t)$ denotes the lapse function and $d\Sigma_k^2$ is the line
element for a D-manifold of constant curvature $k = + 1, 0, - 1$. The
Ricci scalar is given by
\begin{equation}
\label{9}
R=\frac{D}{N^2}\left\{\frac{2\ddot a}{a}+(D-1)\left[\left(\frac{\dot a}{a}
\right)^2 + \frac{N^2k}{a^2}\right]-\frac{2\dot a\dot N}{aN}\right\}.
\end{equation}
Substituting from Eq.(\ref{9}) in the Einstein-Hilbert action for
pure gravity,
\begin{equation}
\label{10}
S_G = \frac{1}{2\kappa} \int d^{(1+D)} x \sqrt{-g}R,
\end{equation}
and using the Hawking-Ellis action of a perfect fluid
for the model universe with variable space dimension the following
Lagrangian has been obtained for decrumpling or TVSD model
(see Ref.\cite{7})
\begin{equation}
\label{11}
L_I := -\frac{V_{D_t}}{2 \kappa N} \left( \frac{a}{a_0} \right)
^{D_t} D_t(D_t-1)
\left[ \left( \frac{\dot a}{a} \right )^2 -\frac{N^2 k}{a^2} \right ]
- \rho N V_{D_t} \left( \frac{a}{a_0} \right )^{D_t},
\end{equation}
where $\kappa=8 \pi {M_P}^{-2}=8 \pi G$, $\rho$ the energy density,
and $V_{D_t}$ the volume of the space-like sections

\begin{eqnarray}
\label{12}
V_{D_t}&=&\frac{2 \pi^{(D_t+1)/2}}{\Gamma[(D_t+1)/2]},\;\;\mbox{closed Universe, $k=+1$,}\\
\label{13}
V_{D_t}&=&\frac{\pi^{(D_t/2)}}{\Gamma(D_t/2+1)}{\chi_c}^{D_t},\;\;\mbox{flat Universe, $k=0$,}\\
\label{14}
V_{D_t}&=&\frac{2\pi^{(D_t/2)}}{\Gamma(D_t/2)}f(\chi_c),\;\;\mbox{open Universe, $k=-1$.}
\end{eqnarray}

Here $\chi_C$ is a cut-off and $f(\chi_c)$ is a function thereof
(see Ref. \cite{7}).

In the limit of constant space dimensions, or $D_t=D_0$,
$L_I$ approaches to the Einstein-Hilbert Lagrangian
which is
\begin{equation}
\label{15}
L_{I}^0 := - \frac{V_{D_0}}{2 \kappa_0 N}
\left( \frac{a}{a_0} \right)^{D_0} D_0(D_0-1)
\left[ \left( \frac{\dot{a}}{a} \right)^2 - \frac{N^2 k}{a^2} \right ]
- \rho N V_{D_0} \left( \frac{a}{a_0} \right )^{D_0},
\end{equation}
where $\kappa_0=8\pi G_0$ and the zero subscript in $G_0$ denotes its
present value. So, Lagrangian $L_I$ cannot abandon Einstein's gravity.
Varying the Lagrangian $L_I$ with respect to $N$ and $a$, we find the
following equations of motion in the gauge $N=1$, respectively
\begin{eqnarray}
\label{16}
&&\left( \frac{\dot a}{a} \right)^2 +\frac{k}{a^2} =
\frac{2 \kappa \rho}{D_t(D_t-1)},\\
\label{17}
&&(D_t-1) \bigg\{ \frac{\ddot{a}}{a} + \left[ \left( \frac{\dot a}{a}
\right)^2
+\frac {k}{a^2} \right] \bigg( -\frac{{D_t}^2}{2C} \frac{d \ln V_{D_t}}{d{D_t}}
-1-\frac{D_t(2D_t-1)}{2C(D_t-1)} 
+\frac{{D_t}^2}{2D_0} \bigg) \bigg\} \nonumber\\
&&+ \kappa p \bigg( -\frac{d \ln V_{D_t}}{d{D_t}} \frac{D_t}{C} 
-\frac{D_t}{C} \ln \frac{a}{a_0} +1 \bigg) =0.
\end{eqnarray}
Using (\ref{5}) and (\ref{16}), the evolution equation of the space
dimension can be obtained by
\begin{equation}
\label{18}
{{\dot{D}}_t}^2= \frac{D_t^4}{C^2} \left[ \frac{2 \kappa \rho}{D_t(D_t-1)}
-k {\delta}^{-2} e^{-2C/{D_t}} \right].
\end{equation}
The continuity equation of decrumpling or TVSD model
can be obtained by (\ref{16}) and (\ref{17})
\begin{equation}
\label{19}
\frac{d}{dt} \left[ \rho \left( \frac{a}{a_0} \right)^{D_t} V_{D_t} \right]
+ p \frac{d}{dt} \left[ \left( \frac{a}{a_0} \right )^{D_t} V_{D_t} \right] =0.
\end{equation}

Let us now study the time variation of the fine structure constant
in decrumpling or TVSD model.
In the International System of Units, the fine structure constant
is given by 
$\alpha \equiv \frac{e^2}{4\pi\epsilon_0 \hbar c} \simeq \frac{1}{137}$
and in the Heaviside-Lorentz System
$\alpha \equiv \frac{e^2}{\hbar c}\simeq \frac{1}{137}$.
It is argued in \cite{8} that in $D$-dimensional spaces the dimensionless
constant of Nature in the Heaviside-Lorentz System is proportional to
$h^{2-D}e^{D-1}G^{\frac{3-D}{2}}c^{D-4}$.
From this for $D=3$ we obtain $\frac{e^2}{h c}$.
The dimensionless quantity $e^2/(h c)$ was first emphasized by Sommerfeld
\cite{9}.

The fine structure constant in $D$-dimensional space is given
by (See Appendix)
\begin{equation}
\label{a20}
{\hat \alpha}= e^{(D-1)}
{\hat k}^{\frac{(D-1)}{2}}
{\hat G}^{(\frac{3-D}{2})}
{\hbar}^{(2-D)}
c^{(D-4)}.
\end{equation}

In decrumpling or TVSD model, the definition of the fine structure
constant in terms of Planck charge is (see Eq.(\ref{22}) in Appendix)
\begin{equation}
\label{a21}
{\hat \alpha} \equiv \left( \frac{e}{{\hat Q}_P} \right)^{(D_t-1)}.
\end{equation}
By using the same approach given in Appendix, one can easily obtain the
fine structure constant in decrumpling or TVSDmodel (see
Eq.(\ref{47}) in Appendix and Eq.(\ref{a20}))
\begin{equation}
\label{a22}
{\hat \alpha}= e^{(D_t-1)}
{\hat k}^{\frac{(D_t-1)}{2}}
{\hat G}^{(\frac{3-D_t}{2})}
{\hbar}^{(2-D_t)}
c^{(D_t-4)}.
\end{equation}
Time derivative of this equation leads to\footnote{We take
the quantities $e$, ${\hat \kappa}$, $\hbar$, and $c$
to be constant.}
\begin{equation}
\label{a23}
\frac{\dot{\hat \alpha}}{\hat \alpha}=
\frac{(3-D_t)}{2}
\frac{\dot{\hat G}}{\hat G} +
{\dot{D_t}} \ln \left( \frac{e}{\hat Q_P} \right).
\end{equation}
Using this equation at the present time and taking
$D_t|_0=3$ we are led to
\begin{eqnarray}
\label{a24}
\frac{\dot{\hat \alpha}}{\hat \alpha} \bigg|_0 &=&
{\dot{D_t}}\bigg|_0 \ln
\left( \frac{e}{\sqrt{4 \pi \epsilon_0 \hbar c}}\right)\nonumber\\
& \simeq & \frac{1}{2}{\dot{D_t}}\bigg|_0 \ln \frac{1}{137}
\end{eqnarray}
From this equation, it can be easily seen that
${\dot {\hat \alpha}}/{\hat \alpha}$ has positive value (${\dot D}_t <0$).
This tells us that the value of the fine structure constant increases
within the cosmic time in TVSD or decrumpling model.
Atomic clocks are one of the principal methods we have to measure
possible variation of the fine structure constant on Earth.
The latest constraint on possible time variation of alpha,
using atomic clocks, is \cite{12} (see also Refs.\cite{{13},{14}})
\begin{equation}
\label{a25}
\bigg| \frac{\dot \alpha}{\alpha} \bigg| \Bigg|_0 < 10^{-15}
{\rm yr}^{-1}.
\end{equation}
Using Eqs.(\ref{a24}) and (\ref{a25}) we are led to
\begin{equation}
\label{a26}
\bigg| {\dot{D_t}} \bigg|_0 < 10^{-15}
{\rm yr}^{-1}.
\end{equation}
This result leads to an upper limit for the absolute value of
the present time variation of the spatial dimension
in decrumpling or TVSD model.

\section*{Appendix}
Let us now obtain a general formula for the fine structure
constant in $D$-dimensional spaces in the International System of
Units. To obtain a general formula for the fine structure constant
in $D$-dimensional spaces we use the definition of the fine structure
constant in terms of Planck charge $Q_P$
\begin{equation}
\label{20}
Q_P \equiv \sqrt{ 4 \pi \epsilon_0 \hbar c}.
\end{equation}
In $3$-dimensional spaces, the definition of the fine structure constant
in terms of Planck charge is
\begin{equation}
\label{21}
\alpha \equiv \left( \frac{e}{Q_P} \right)^2.
\end{equation}
In $D$-dimensional spaces, a generalized form of (\ref{21})
defines the fine stucture constant. Defining
$\hat \alpha$ as the fine structure constant in $D$-dimensional
spaces, we have
\begin{equation}
\label{22}
{\hat \alpha} \equiv \left( \frac{e}{{\hat Q}_P} \right)^{(D-1)},
\end{equation}
where ${\hat Q}_P$ is Planck charge in $D$-dimensional spaces
and as a function of $D$.
To obtain the fine stucture constant in higher dimensions, we must
first obtain Planck charge in higher dimensions.
In doing so, we propose
that in $D$-dimensional spaces, Planck charge can be written in
terms of four foundamental constants
$${\hat k},\;\;\;{\hat G},\;\;\;\hbar,\;\;\;c,$$
where ${\hat k}$ is the electrostatic coupling constant
and ${\hat G}$ is the gravitational constant in
$D$-dimensional spaces. We also define four unknown functions of the
space dimension 
$$\beta(D),\;\;\;\eta(D),\;\;\;\xi(D),\;\;\;\tau(D),$$
so that these four unknown functions of $D$ are the exponents
of ${\hat k}$, ${\hat G}$, $\hbar$ and $c$ in the formula of
Planck charge 
\begin{equation}
\label{23}
{\hat Q}_P={\hat k}^{\beta(D)}{\hat G}^{\eta(D)}\hbar^{\xi(D)}c^{\tau(D)}.
\end{equation}
Coulomb's law for the electrostatic force in $D$-space and $3$-space
are defined by
\begin{eqnarray}
\label{24}
F_D &=& \frac{{\hat k} q_1 q_2}{r^{D-1}},\\
\label{25}
F &=&\frac{k q_1 q_2}{r^2},
\end{eqnarray}
where $k=1/(4 \pi \epsilon_0)$.
Newton's law for the gravitational force in $D$-space and $3$-space
are defined by
\begin{eqnarray}
\label{26}
F_D &=& \frac{{\hat G} m_1 m_2}{r^{D-1}},\\
\label{27}
F &=& \frac{G m_1 m_2}{r^2}.
\end{eqnarray}
Using Gauss law in $D$-dimensional spaces, one can derive
(see Ref. \cite{10} for more detailed explanation)
\begin{equation}
\label{28}
G = \frac{S_D}{4 \pi} \frac{\hat G}{V_{(D-3)}},
\end{equation}
and
\begin{equation}
\label{29}
k=
\frac{S_D}{4 \pi} \frac{\hat k}{V_{(D-3)}},
\end{equation}
where
\begin{equation}
\label{30}
S_D \equiv \frac{2 \pi^{D/2}}{\Gamma \left( \frac{D}{2} \right)},
\end{equation}
is the surface area of the unit sphere in $D$-dimensional spaces and
$V_{(D-3)}$ is the volume of $(D-3)$ extra spatial dimensions.
For $D=3$, we have $S_3=4 \pi$.
The units of ${\hat k}$ and ${\hat G}$ can be easily found by
(\ref{29}) and (\ref{28}), respectively. Moreover, in $D$-space the
units of $\hbar$ and $c$ is the same as their units in $3$-space.
So, in terms of mass $(M)$, length $(L)$, time $(T)$ and electric
charge $(Q)$, in $D$-dimensional spaces, we have
the following units for $\hat k$, $\hat G$, $\hbar$ and $c$
\begin{eqnarray}
\label{31}
\left[ {\hat k} \right] &=& Q^{-2} M L^{D} T^{-2},\\
\label{32}
\left[ {\hat G} \right] &=& M^{-1} L^D T^{-2},\\
\label{33}
\left[ \hbar \right] &=& M L^2 T^{-1},\\
\label{34}
\left[ c \right] &=& L T^{-1}.
\end{eqnarray}
We also know that the units of Planck charge ${\hat Q}_P$
is equal to the unit of electric charge, $Q$,
\begin{equation}
\label{35}
\left[ {\hat Q}_P \right]=Q.
\end{equation}
Therefore, we can rewrite (\ref{23}) in terms of units
\begin{eqnarray}
\label{36}
Q =
\left( Q^{-2} M L^{D} T^{-2} \right)^{\beta(D)}
\left( M^{-1} L^D T^{-2} \right)^{\eta(D)}
\left( M L^2 T^{-1} \right)^{\xi(D)}
\left(L T^{-1} \right)^{\tau(D)}.
\end{eqnarray}
To satisfy (\ref{36}) for the units of Planck
charge in $D$-dimensional spaces,
in the right-hand side of (\ref{36}) the exponents of $M$
must be vanished
\begin{equation}
\label{37}
\beta - \eta + \xi =0,
\end{equation}
the exponents of $L$ must be vanished 
\begin{equation}
\label{38}
D \beta + D \eta + 2 \xi + \tau =0,
\end{equation}
the exponents of $T$ must be vanished 
\begin{equation}
\label{39}
2 \beta + 2 \eta + \xi + \tau=0,
\end{equation}
and finally the exponents of $Q$ must be equal to one
\begin{equation}
\label{40}
-2 \beta =1.
\end{equation}
Using four equations (\ref{37})-(\ref{40}) we can obtain four
functions $\beta$, $\eta$, $\xi$ and $\tau$ with respect to $D$
\begin{eqnarray}
\label{41}
\beta(D) &=& - \frac{1}{2},\\
\label{42}
\eta(D) &=& \frac{D-3}{2(D-1)},\\
\label{43}
\xi(D) &=& \frac{D-2}{D-1},\\
\label{44}
\tau(D) &=& \frac{4-D}{D-1}.
\end{eqnarray}
Substituting (\ref{41})-(\ref{44}) in (\ref{23}) we obtain Planck charge
as a function of $D$ in $D$-dimensional spaces
\begin{equation}
\label{45}
{\hat Q}_P=
{\hat k}^{-\frac{1}{2}}
{\hat G}^{\frac{D-3}{2(D-1)}}
{\hbar}^{\frac{D-2}{D-1}}
c^{\frac{4-D}{D-1}}.
\end{equation}
Now, substituting (\ref{45}) in (\ref{22}) yields the fine structure
constant in $D$-dimensional spaces
\begin{eqnarray}
\label{46}
{\hat \alpha} &=& \left( \frac{e}{{\hat Q}_P} \right)^{(D-1)}
= \left( \frac{e}
{ {\hat k}^{-\frac{1}{2}}
{\hat G}^{\frac{D-3}{2(D-1)}}
{\hbar}^{\frac{D-2}{D-1}}
c^{\frac{4-D}{D-1}}
} \right)^{(D-1)},
\end{eqnarray}
or
\begin{equation}
\label{47}
{\hat \alpha}= e^{(D-1)}
{\hat k}^{\frac{(D-1)}{2}}
{\hat G}^{(\frac{3-D}{2})}
{\hbar}^{(2-D)}
c^{(D-4)}.
\end{equation}
Eq.(\ref{47}) gives us the fine structure constant in $D$-dimensional
spaces and in the International System of Units and reads
$\alpha=\frac{e^2}{4 \pi \epsilon_0 \hbar c}\simeq \frac{1}{137}$ in
$3$-dimensional spaces. It is worth mentioning that in Ref.\cite{11} by using another approach
the fine structure constant in $D$-dimensional space has been obtained.
Our result here, i.e. Eq.(\ref{47}), is in agreement with the result
presented in Ref.\cite{11}.

\end{document}